# Quantized orbital angular momentums of dipolar magnons and magnetoelectric cavity polaritons

E. O. Kamenetskii

School of Electrical and Computer Engineering, Ben Gurion University of the Negev, Israel

June 17, 2024

**Abstract**
Magnons are viewed as local deviations from the ordered state. Usually, the spin magnetic moment of magnons is considered. In a 3D-confined structure of a magnetic insulator with magnetodipolar mode (MDM) oscillations, an orbital angular momentum (OAM) as well as a spin angular momentum (SAM) can be observed along a static magnetic field. In such a confined structure as quasi-2D ferrite disk, energy levels of MDM oscillations are quantized. Quantum confinement is characterized by a half-integer internal OAM, which is also associated with a circulating energy flow. The observation of MDM resonances in the 3D-confined structure of a magnetic insulator is due to the interaction of two subsystems: ferromagnetic and electric polarization orders. The coupling states of these two concurrent orders, caused by OAMs, are considered as magnetoelectric (ME) states. The fields in the vicinity of MDM resonators are characterized by simultaneous violation of time reversal and inversion symmetry. This plays a significant role in the problems of strong light-matter interaction regime and quantum atmosphere. The analysis of SAM and OAM in 3D-confined magnetic insulators becomes very important for the realization of ME meta-atomic structures. Current interest lies in considering such artificial systems as subwavelength ME quantum emitters of electromagnetic radiation.

## I. INTRODUCTION

Magnons are spin-wave excitations in insulating magnetic materials [1]. Can magnon excitations have both the SAM, $S$, and OAM, $\mathcal{L}$? This question has been arising for over a century, since the experiments of Einstein and de Hass [2] and Barnett [3]. Whereas a single spin flip of magnon is $S = \pm\hbar$, the OAM, $\mathcal{L}$, of such a magnon is unknown.

The OAM of magnons is considered as a complementing momentum brought about by spin-orbit coupling. In noncollinear spin states with Dzyaloshinskii-Moriya interaction, spin magnetization and orbital magnetization are shown as independent quantities [4]. The OAM of magnons in collinear ferromagnet and antiferromagnetic systems with nontrivial networks of exchange interactions are studied in Ref. [5]. Since the total angular momentum $I = S + \mathcal{L}$ should be conserved, magnons in a confined magnetodielectric will be sensitive to external fields. The OAMs in nanocylinder nanotube structures were analyzed in Refs. [6 – 8]. Based on the effects of Brillouin light scattering in magneto-optical structures, the OAM conservation of circulating magnons in ferromagnetic sphere and whispering gallery mode cavity were studied in Refs. [9 – 12]. OAMs in magnon beams were considered in Ref. [13].

The role of SAM and OAM in 3D-confined magnetic insulators becomes very important for the realization of meta-atomic structures. Current interest lies in considering such artificial systems as subwavelength quantum emitters of electromagnetic radiation. The intrinsic dynamical processes in meta-atoms must be distinguished by the presence of energy eigenstates with SAM and OAM degrees of freedom. In recent studies [14], quantum confinement of MDM oscillations (known also as Walker's magnetostatic-magnon oscillations [1]) has been observed in a quasi-2D ferrite disk placed inside a microwave structure. An analysis shows that an OAM of MDMs in a ferrite disk is a half-integer internal angular momentum related to a circulating flow of energy. The observation of MDM resonances in the 3D-confined structure of a magnetic insulator is due to the interaction of two subsystems: ferromagnetic and electric polarization orders. The coupling states of these two concurrent orders, caused by OAMs, are considered as ME states. The symmetry broken field patterns with unique properties (so-called ME fields [15 – 17]) are observed in a vacuum vicinity zone of a ME sample. The conservation of total angular momentum follows from invariance under rotations of fields keeping the solid fixed. This is the conservation of pseudoangular momentum, which is different from the conservation of angular momentum following from invariance under rotations of all the constituents of a solid [18]. On the other hand, the presence of effective masses and effective moment of inertia of MDMs in a quasi-2D ferrite disk [19, 20] raises the question on mutual optomechanical interaction mediated by the radiation-pressure force [21].

The symmetry-breaking properties of dipolar magnons leave a characteristic imprint in virtual photons of quantum vacuum fluctuations near a 3D-confined magnetic structure. In particular, the angular momentum of virtual photons can induce the Casimir torque. A key question is whether these signatures would be strong enough to observe. Probing quantum atmospheres might offer a way to identify new types of symmetry breaking. A special interest is to consider quantum atmosphere from axion electrodynamics [22]. Since the quantum atmosphere is influenced by external fields, an analysis suggests how to probe symmetry-breaking atmospheres experimentally.

In this paper, we consider the intrinsic magnon spin-orbit coupling in a 3D-confined magnetic insulators due to dipolar interaction. The MDM-polariton model for ME quantum emitters is analyzed. In the fields scattered by a ME particle, one should distinguish between left and right. In ME polaritons in a microwave cavity, unilateral circulations of power flows occur due to the curvature of the wave fronts. These are helical modes in vacuum. For each of these modes, originated from MDM resonances, there is a specific topology.

## II. OAMS OF MAGNONS AND *PT* SYMMETRY

### A. *G* and *L* modes of MDM oscillations in quasi-2D ferrite disks

The effect of OAM in 3D-confined magnetic insulators can be observed for MDM [or magnetostatic (MS)] oscillations. These are long-wavelength modes of oscillations, where exchange and electromagnetic propagation in a ferromagnetic sample are simultaneously ignored [23]. The fundamental role of the orbital response in the angular momentum dynamics presumes the specific orbital circulations of magnetization. Specific properties of spin wave propagation in confined geometries of magnetic isolators are associated with the dependence of magnon energy on a bias magnetic field. The magnon energy in 3D-confined ferrite sample can be quantized. OAMs of magnons in such samples are quantized as well. These statements, important when trying to realize coherent EM manipulation of magnons in the quantum regime, become obvious when correctly solving the boundary value problem for MDM spectra in a ferromagnetic sample. Along a boundary of a sample, orbital currents of magnetic dipoles occur. Such surface magnon currents carry a fixed amount of orbital energy and orbital momentum and enables electric manipulation of magnetization [14, 19].

When considering symmetries and their corresponding conservation laws, it is necessary to correctly take into account the boundary conditions. As is known, in solving the boundary value problem that involves the eigenfunctions of a differential operator, the boundary conditions must be in a definite correlation with the type of this differential operator. Surprisingly, this fact, well known from textbooks of functional analysis [24, 25], is often not taken into consideration in numerous publications regarding magnetostatic oscillations in ferrite samples. In spectral solutions obtained based on the Walker's second-order differential equation for magnetostatic wave functions, the boundary conditions of continuity of the normal component of magnetic induction and tangential component of magnetic field are accompanied by the boundary conditions for dynamic magnetization [6, 7, 10, 26]. But the boundary conditions for dynamic magnetization cannot be considered as suitable boundary conditions for the Walker-equation dipolar magnons. The orthogonality relations derived for dynamic magnetization of long-wavelength magnons [23, 26, 27] are not appropriate for selfadjointness of the Walker-equation differential operator. In this case, the eigenfunctions do not form a complete set of spectral solutions.

To make the MDM spectral problem analytically integrable, two approaches based on the use of MS-potential wave function $\psi(\vec{r},t)$ have been proposed. These approaches, distinguished by differential operators and boundary conditions, give two types of MDM oscillation spectra in a quasi-2D ferrite disk. The approaches are conditionally called the *G* and *L* modes in the magnetic dipolar spectra [14, 19, 20, 28, 29]. The MS-potential wave function $\psi(\vec{r},t)$ manifests itself in different manners for each of these types of these spectra.

The boundary value problem is solved based on Walker equation [1, 23]

$$\vec{\nabla} \cdot \left( \vec{\vec{\mu}} \cdot \vec{\nabla} \psi \right) = 0 \tag{1}$$

inside and the Laplace equation $\nabla^2 \psi = 0$ outside a ferrite sample. Here $\vec{\vec{\mu}}(\omega, \vec{H}_0)$ is a tensor of ferrite permeability at the ferromagnetic-resonance frequency range. For a ferrite magnetized along *z* axis, the permeability tensor has a form [1]:

$$\vec{\vec{\mu}} = \mu_0 \begin{bmatrix} \mu & i\mu_a & 0 \\ -i\mu_a & \mu & 0 \\ 0 & 0 & 1 \end{bmatrix}. \tag{2}$$

In a quasi-2D ferrite disk with the disk axis oriented along *z*, the *L*-mode solution for the MS-potential wave function is written in a cylindrical coordinate system as:

$$\psi = C\xi(z)\tilde{\varphi}(r,\theta), \tag{3}$$

where $\tilde{\varphi}$ is a dimensionless membrane function, $r$ and $\theta$ are in-plane coordinates, $\xi(z)$ is a dimensionless function of the MS-potential distribution along $z$ axis, and $C$ is a dimensional amplitude coefficient. Being the energy-eigenstate oscillations, the MDMs in a ferrite disk are also characterized by topologically distinct structures of the fields. This becomes evident from the boundary condition on a lateral surface of a ferrite disk of radius $\mathcal{R}$, written for a membrane wave function as:

$$\left(\frac{\partial \tilde{\varphi}}{\partial r}\right)_{r=\mathcal{R}^+} - \mu\left(\frac{\partial \tilde{\varphi}}{\partial r}\right)_{r=\mathcal{R}^-} = i\frac{\mu_a}{\mathcal{R}}\left(\frac{\partial \tilde{\varphi}}{\partial \theta}\right)_{r=\mathcal{R}^-}. \tag{4}$$

Evidently, in the solutions, one can distinguish the time direction (given by the direction of the magnetization precession and correlated with a sign of $\mu_a$) and the azimuth rotation direction (given by a sign of $\frac{\partial \tilde{\varphi}}{\partial \theta}$). For a given sign of a parameter $\mu_a$, there can be different MS-potential wave functions, $\tilde{\varphi}^{(+)}$ and $\tilde{\varphi}^{(-)}$, corresponding to the positive and negative directions of the phase variations with respect to a given direction of azimuth coordinates. So, a function $\tilde{\varphi}$ is not a single-valued function. It changes a sign when angle $\theta$ is turned on $2\pi$. The term in the right-hand side of Eq (4) is considered as a topological magnetic current. The spectral theory developed based on orthogonal single-valued membrane functions and topological magnetic currents shows the ME effect from a viewpoint of the Berry phase connection [19].

For the known solution of the wave function $\psi(\vec{r},t)$, the power flow density is viewed as a current density [19]:

$$\vec{\mathcal{J}} = \frac{i\omega}{4}\left(\psi\left(\vec{\mu}\cdot\vec{\nabla}\psi\right)^* - \psi^*\left(\vec{\mu}\cdot\vec{\nabla}\psi\right)\right) \tag{5}$$

This is a power flow arising from the dipole-dipole interaction of magnetic dipoles. If MDMs in a ferrite disk are unidirectionally rotating (chiral) waves, circulation of vector $\vec{\mathcal{J}}$ along contour $\mathcal{C} = 2\pi r$ is not equal to zero. In this case, we should observe an OAM due to the power-flow circulation:

$$\vec{\mathcal{L}}_z = \oint_\mathcal{C} \vec{r}\times\vec{\mathcal{J}}\,dl. \tag{6}$$

However, the questions of whether such a closed-loop circulation of power flow is possible and whether it can be quantized are nontrivial.

For the *G*-mode solution, the boundary condition on a lateral surface is similar to Eq. (4), but with a zero right-hand side. So, a membrane function for the *G*-mode is a single-valued function. In the case of the *G*-mode spectrum, where the physically observable quantities are energy eigenstates, the MS-potential wave functions appear as Hilbert-space scalar wave functions. For *G* modes, one can formulate the energy eigenstate boundary problem based on the Schrödinger-like equation for scalar-wave eigenfunctions $\psi(\vec{r},t)$ with using the Dirichlet-Neumann (ND) boundary conditions. However, in the MDM spectral problem, the ND boundary conditions are not the EM boundary conditions. While the considered above ND boundary conditions are the so-called essential boundary conditions, the EM boundary conditions, used for *L* modes, are the natural boundary conditions [24]. In the case of the *L* modes, the MS-potential wave functions are considered as generating functions for the vector harmonics of the magnetic and electric fields. In a quasi-2D ferrite disk placed in a microwave cavity, one observes quantum confinement effects of MDM oscillations. These modes, characterized by energy eigenstates with rotational superflows and quantized vortices, are exhibited as spinor condensates. Along with the condensation of MDM magnons in the quasi-2D disk of the magnetic insulator, electric dipole condensation is also observed. At the MDM resonances, transfer between angular momenta in the magnetic insulator and in the vacuum cavity, demonstrates generation of vortex flows with fixed handedness. This indicates unique topological properties of polariton wavefronts [14].

**B. The OAM balance of MDM oscillations in a system of two coupled ferrite resonators**

OAMs in 3D-confined magnetic insulators are closely related to *PT* symmetry effects. *PT*-symmetry is invariance with respect to combined space reflection *P* and time reversal *T*. Maxwell's equations are time-reversal invariant. At the same time, EM wave propagation in gyrotropic media is related to violation of time-reversal symmetry. This effect is associated with a given direction of a bias magnetic field. For the opposite direction of bias magnetic field, there is a transformation $t \rightarrow -t$. In a lossless magnetic insulator structure, the time reversal operation is actually a magnetization reversal. Due to invariance of the motion equations of microscopic currents with respect to sign of time, the total current is zero. So, the thermodynamics equilibrium state of magnetic medium does not change at the time reversal operation [30]. The only fundamental theory that emphasizes the preferred direction of time is the second law of thermodynamics, which states that entropy increases as time flows into the future. It provides the orientation of time, which can be called, in other words, the arrow of time. This is a thermodynamic arrow. When dissipation or the overall increase in entropy occurs, the orientation of the arrow of time is obvious. On the other hand, in a lossless structure, radiation suggests increased entropy. So, the radiative arrow is linked to the thermodynamic arrow. EM waves transfer information. They are radiative outwards from their sources. The real power flow can be observed in a lossless waveguide only due to the *PT*-symmetry. This means that EM propagation is a lossless waveguide, either non-gyrotropic or gyrotropic, is only possible if backward propagation is also assumed. In this case, thermodynamics equilibrium state of a waveguide does not change at the time reversal operation.

Gyrotropic media with nonsymmetrical constitutive tensors caused by an applied DC magnetic field have been called "nonreciprocal" media because the usual reciprocity theorem does not apply to them. Rumsey has introduced a quantity called the "reaction" and interpreted it as a "physical observable" [31]. This made it possible to obtain a modified reciprocity theorem based on the property of gyrotropic media that nonsymmetrical constitutive tensors of permittivity or permeability are transposed by reversing the DC magnetic field [32]. In this regard, it is worth noting that the quadratic relations for fields in the analysis of nonreciprocity (Lorentz's theorem) differ from the quadratic relations for fields in the analysis of the balance of energy (Poynting's theorem). The scattering matrix for a reciprocal network is symmetric, while the scattering matrix for a lossless network is unitary [1, 33]. Recent publications show that in a lossless gyrotropic structure with broken time reversal symmetry, a "one-way waveguide" can be realized [34 – 36]. It is assumed that this is a channel though which electromagnetic energy can propagate in only a single direction. However, the one-way direction of power flow is related to the time direction in which entropy increases.

Suppose that we make a ring from such a one-way waveguide. In this waveguide ring, a small local perturbation of the field amplitude will lead to the system being taken out of thermodynamic equilibrium since the power flow can pass only in one azimuthal direction. In these states, the field amplitude will grow as time increases. So, we need a mechanism that suppresses anti-resonant states, making the resonant states dominant for positive times. It means that according to the second law of thermodynamics, it is impossible to implement a one-sided waveguide without losses. This fact is well illustrated by the example of a microwave ferrite isolator. This is a two-port non-reciprocal device. Its scattering matrix indicates that the power flow is transmitted in only one direction. However, this scattering matrix is *not unitary*. This means that the isolator – the "one-way waveguide" – must be lossy [1, 33]. The above analysis is closely related to the problem of OAM of magnons. Importantly, "one-way waveguide" ring can be realized in lossless gyrotropic structure with the *PT* symmetry. For modes in such a structure we have a spectrum of a non-Hermitian Hamiltonian admitting a complete set of biorthonormal eigenvectors.

In the case of the MDM resonator, we have a chosen direction of time. Let us consider the dynamics of magnetization on any orbit in a 3D-confined structure. The orbit is on the *xy* plane. The center of this orbit lies on the symmetrical axis of the sample, pointed along the *z* axis. The bias magnetic field is directed along or opposite to the *z*-axis: there are the $+\vec{H}_0$ or $-\vec{H}_0$ bias magnetic fields (see. Fig. 1). Transitions in time are counted by a Larmor clock. The Larmor clock is commonly used to measure the duration of an event by means of the *precession of an electron spin* in a homogeneous magnetic field [37]. We use the Larmor clock concept to show the classic phase times.

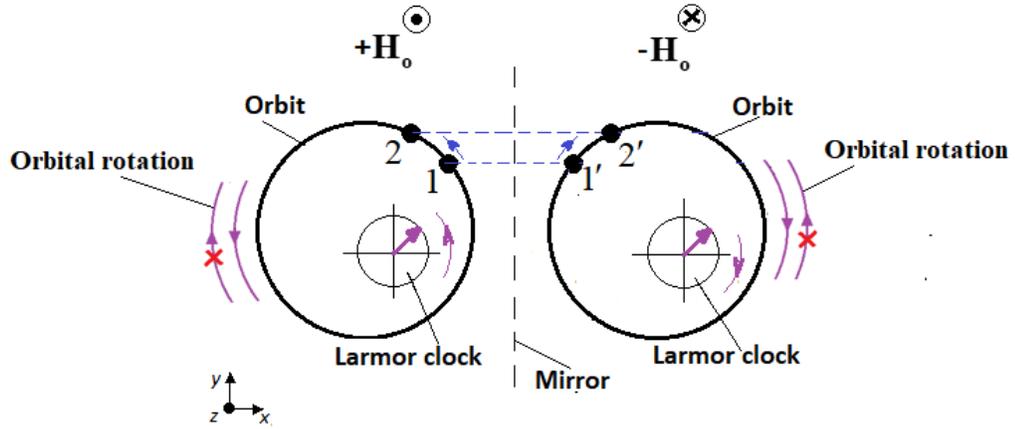

**Fig. 1.** The CCW and CW orbital rotations at the $+\vec{H}_0$ or $-\vec{H}_0$ bias magnetic fields. Transitions in time are counted by a Larmor clock.

For a bias magnetic field $+\vec{H}_0$, we can have only a counterclockwise (CCW) orbital rotation. At a bias magnetic field $-\vec{H}_0$ we can have only a clockwise (CW) orbital rotation. Suppose now that at $+\vec{H}_0$, we shift a process from an initial state 1 on the orbit to a new state 2, that is we make the transition $1 \rightarrow 2$. We then strive to return to original state 1. We cannot return along the shortest path to the same place by rotating CCW on a circle. When we assume that at $+\vec{H}_0$, we have the transition on the orbit from an initial state at the azimuth angle $\theta = 0$ to the same state at the azimuth angle $\theta = 2\pi$, we will not have a *quantized* state. The reason is that for such a $\theta = 2\pi$ rotation we cannot generate a *back-action*, necessary for any resonance (*normalized*) state. Out of equilibrium, the action-reaction symmetry of the interactions is broken. The back-action can be generated if we *additionally* use another disk which has a $\theta = 2\pi$ rotation at $-\vec{H}_0$ bias magnetic field. Due to changes in both the direction of the bias magnetic field and the direction of an orbital rotation, we observe the break of both spatial and time symmetries. In such a *PT* symmetric system, MDM resonators, connected through long-range dipolar interaction, are in a *biorthogonality* relationship.

In any separate disk, the *resonant state* occurs at the $4\pi$ orbital rotation. At this the $4\pi$ - orbital rotation, the Larmor clocks, both at the $+\vec{H}_0$ or $-\vec{H}_0$ bias magnetic fields, are at the $2\pi$ spin rotations. The orbital rotation is synchronized with the spin rotation. It other words, the orbital motion is correlated with the Larmor clock. The Larmor clock is associated with the motion of magnetization with frequency $\omega_L$. While at the frequency of the spin rotation $\omega = \omega_L$ we have the SAM $S = \pm\hbar$, the orbital rotation occurs at the frequency $2\omega_L = 2\omega$. It means that the OAM of magnon is $\mathcal{L} = \pm\frac{1}{2}\hbar$. Why we have only CCW orbital rotation at the field $+\vec{H}_0$ and only a clockwise CW orbital rotation at the field $-\vec{H}_0$? This question can be answered by analyzing the spectral problem for MDM oscillations in a quasi-2D ferrite disk [14, 19, 28, 29, 38].

The *PT*-symmetric system of two ferrite disks placed in a microwave cavity is shown in Fig. 2. Assuming there is OAM on each disk due to the power-flow circulation [see Eq. (6)], we have synchronized orbital rotations with power-flow vortices in a system of two disks. This is illustrated in Fig. 3.

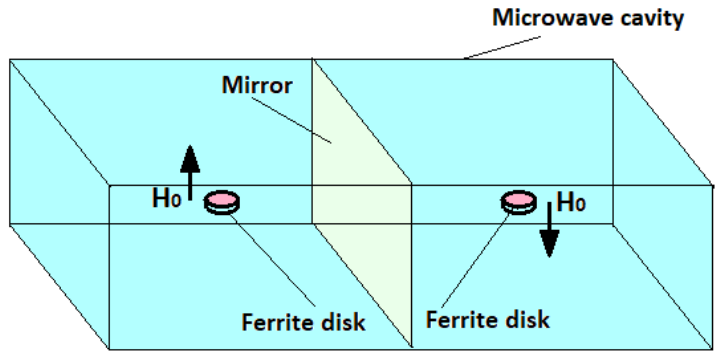

**Fig. 2.** *PT*-symmetric system of coupled MDM resonators.

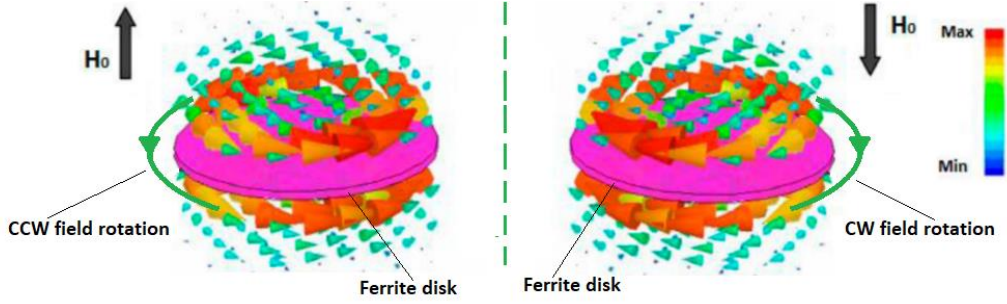

**Fig. 3.** Power-flow vortices of two ferrite-disk resonators coupled due to long-range dipolar interaction. OAM balance occurs when both types of magnetic bias fields are used.

As we discussed above, unilateral propagation of power flow in a closed loop is impossible from the point of view of thermodynamics. In other words, it is impossible to observe the one-way closed-loop circulation of the arrow of time. Nevertheless, in a system of two coupled MDM resonators, unilateral circulations of power flows can be assumed. To clarify this statement, we initially dwell on *PT* symmetry in a gain/loss structure. In this system with a non-Hermitian Hamiltonian, one has time evolution of amplitudes with amplification and dissipation. When, due to dissipation or amplification respectively, we increase or decrease entropy, the direction of the arrow of time becomes obvious. Fig. 4 (*a*) illustrates this situation in a system with an antisymmetric gain/loss profile. Let us now assume that there is a structure of two MDM resonators in which CCW and CW closed-loop circulations of power flow occur. In such a structure with CCW and CW closed-loop circulations of the arrows of time, one can observe the increase and decrease in entropy. The thermodynamic state of one resonator can be changed only at the expense of effects on another resonator. To achieve thermodynamic equilibrium, the entire system of two resonators must be in a state where one of the MDM resonators (for example, resonator with $+\vec{H}_0$ bias magnetic field) does *positive work* (leading to an increase in entropy) and the other resonator (resonator with $-\vec{H}_0$ bias magnetic field) does *negative work* (which leads to a decrease in entropy). To realize this condition, the closed-loop unilateral wave propagations in the resonators must be mutually synchronized. This synchronization occurs due to Larmor clocks, operating in both resonators. This effect is illustrated in Fig. 4 (*b*).

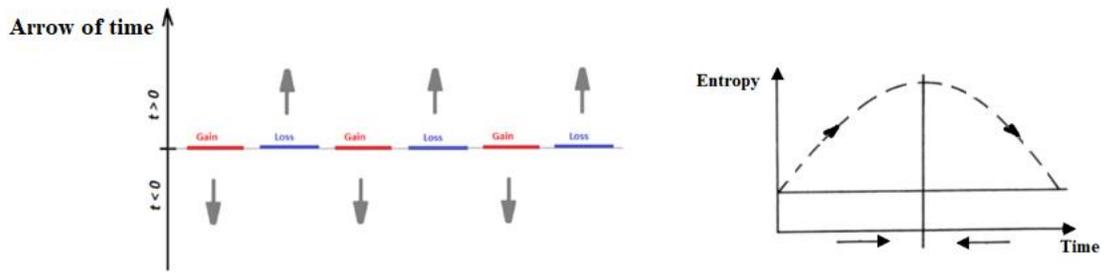

(*a*)

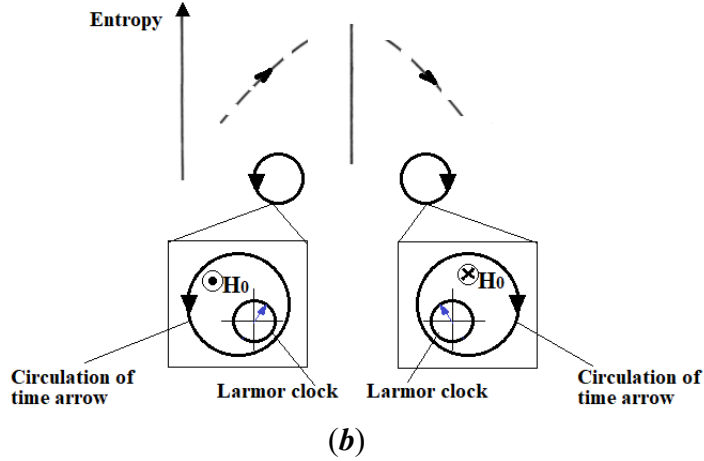

(b)

**Fig. 4.** One-way direction of power flow is associated with one-way direction of time. This is related to the increase or decrease in entropy. (*a*) Arrows of time; (*b*) circulations of arrow of time.

## III. MDM FERRITE DISK IN A MICROWAVE CAVITY

In the *PT*-symmetric structure discussed above, there is a balance of angular momentum due to biorthogonal relationships in the system of two coupled ferrite resonators. At the same time, in microwave experiments with magnon oscillations [39 – 42], a *single* ferrite disk is used in a microwave structure. To understand the underlying mechanism of angular momentum balance in such a single-disk system, it is necessary to analyze how a MDM ferrite disk interacts with metal walls of the microwave waveguide. In these dynamic processes in magnetic insulators, it is necessary to take into account not only magnetostatic (MS) interactions, but also electrostatic (ES) interactions.

In a quasi-2D ferrites disk, the dynamical electric polarization is induced at MDM resonances due to a topological effect. On the other hand, electric polarization currents generate a topological contribution to dynamical magnetization [43]. The observation of resonances in the confined structure of a magnetic insulator is mainly due to the interaction of two subsystems: ferromagnetic and electric polarization orders. The coupling states of two, MS and ES, concurrent orders are considered as ME states. The fields in the vicinity of these resonators, called the ME near fields, are characterized by simultaneous violation of time reversal and inversion symmetry. Quantum confinement effects are considered based on the concept that the MS function $\psi$ ($\vec{H} = -\vec{\nabla}\psi$) and the ES function $\phi$ ($\vec{E} = -\vec{\nabla}\phi$) are *scalar wave functions*. Each of the scalar wave functions is characterized by a module and a phase. In spectral problem solutions, MS and ES wave functions play the role of order parameters for interacting magnetic and electric dipoles. The mesoscopic-scale quantization is described based on the MS and ES wave functions. Such quantized states are well observed in the cavity ME effect.

Due to the topological action of the azimuthally unidirectional transport of energy in a MDM-resonance ferrite sample there exists the opposite topological reaction on a metal screen placed near this sample. We call this effect topological Lenz's effect [44]. The topological Lenz's law is applied to opposite topological charges: one in a ferrite sample and another on a metal screen. The MDM-originated near fields – the ME fields – induce helical surface electric currents and effective topological charges on the metal. The fields formed by these currents and charges will oppose their cause. This OAM balance condition is illustrated in Fig. 5.

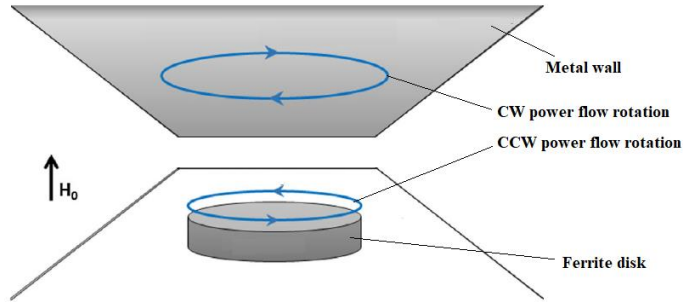

Fig. 5. The OAM balance condition: Along the direction of the bias magnetic field, there are opposite power flow circulations on the surface of a ferrite disk and on the metal surface.

In Fig. 6, we show a structure in which a small ferrite disk is placed in a rectangular waveguide in close proximity to the waveguide walls. It is supposed that for the main $TE_{10}$ mode of microwave waveguide, there are no EM retardation effects in vacuum cylinders above and below a ferrite disk. Fig. 6 represents the situation where CCW rotation of the electric fields in a ferrite disk is in balance with CW rotation of charges and currents induced on the metal walls. The inserts in Fig. 6, obtained based on numerical results in Ref. [44], indicate the existence of "topological charges" at the MDM resonances. The insertion in Fig. 6 shows the topology of surface electric currents on the upper wall of a waveguide. The whole picture of this topological structure rotates clockwise [44]. In quasi-2D ferrite disk and in vacuum cylinders above and below a ferrite disk, we have quasistatic solutions for the *helical* fields. Such helical fields arise from the spectral solution of MS and ES wave functions in a *helical coordinate system* [14, 38]. The helical-mode quasistatic resonances can occur [38].

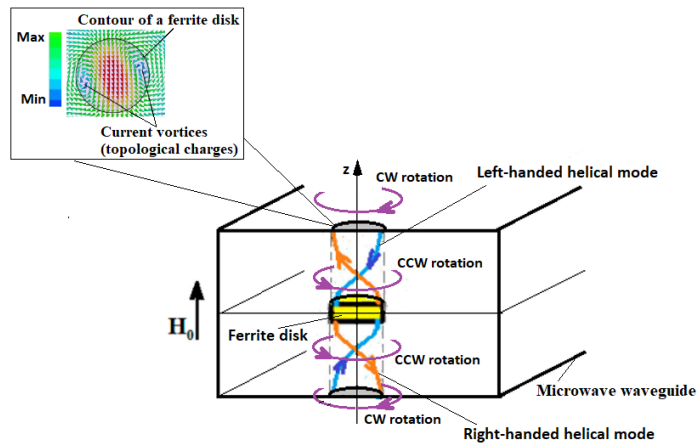

**Fig. 6.** Quasistatic solutions for helical modes. Rotation of the electric field in a ferrite disk is in the angular momentum balance with the rotation of electric currents and topological charges induced on the metal walls. We can observe *braided fields in vacuum* cylinders above and below a ferrite disk. The insertion shows the topology of surface electric currents on the upper wall of a waveguide. The whole picture of this topological structure rotates clockwise.

In Fig. 6, one can see the right-handed and left-handed helical modes of the electric fields in the vacuum cylinders. The role of the vacuum cylinders above and below a ferrite disk is crucial. In the absence of vacuum gaps between the ferrite and metallic surfaces, the MDM spectrum collapses. The effect of Larmor-clock mechanism is manifested on the metal walls. We have circulations of arrow of time both in the ferrite disk and on the metal walls. Curved field lines (helical lines) are thermodynamically favored.

Inside a ferrite disk, the total angular momenta of CW and CCW modes do not eliminate each other. The CW – CCW asymmetry in a ferrite disk would give the vacuum an angular momentum in one azimuth direction, and the metal wall of the cavity would have to gain momentum in the opposite direction to compensate. The spin and orbital angular momentum of vacuum zero fluctuations contribute to the motion of electrons on the metal walls of the cavity. The possibility of extracting momentum from vacuum was discussed by Feigel [45]. Feigel argued that the momentum of vacuum zero fluctuations can occur only in a structure with *PT*-symmetry breakings. The main idea is to suggest a new quantum mechanical effect, namely the extraction of momentum from the electromagnetic vacuum oscillations. In the proposed effect, *momentum* is extracted from a vacuum field. This is different from the case of the Casimir effect, in which *energy* is extracted from a vacuum field [46]. In Ref. [47], Feigel argued that rotating ME particles can generate changes in momentum of zero-point fluctuations, which result in the self-propulsion in quantum vacuum. The self-propulsion requires mechanical back-action from ME particle. To provide this ME particle should be a propellor-like device. In our case, we see that, while keeping the solid body motionless, the rotating fields of ME particles modify the momentum of quantum vacuum.

Returning to the experiments [39 – 42], it should be noted that in the microwave structures considered in these experiments, the MDM ferrite disk is not placed in close proximity to the metal walls of the waveguide. In a structure where the ferrite disk is located at large distances from the walls of the cavity, it is possible to analyze the oscillation spectra of the MDM resonances in *cylindrical coordinate system*. In the magnetic subsystem of the ferrite-disk resonator, localized distribution of an edge magnetic current is viewed as an *electric flux*. For a magnetic current loop, we have an electric moment. Two anti-parallel electric dipoles are visualized as a quadrupole moment. Localized distribution of an edge electric current in the electric subsystem is viewed as a *magnetic flux*. At a large distance we observe a magnetic dipole due to the electric current loop [43].

In the structures [39 – 42], the EM retardation effects should be taken into account. We will classify the interaction of cavity photons with the MDM ferrite disk as *ME cavity polaritons*. When observing a MDM ferrite disk in a cavity, it is necessary to use a description of the spectral response functions of the system with respect to two external parameters – a bias magnetic field $H_0$ and a signal frequency $\omega$ – and analyze the correlations between the spectral response functions at different values of these external parameters. The spectra of MDM oscillations in a quasi-2D ferrite disk are shown schematically in Fig. 7. A very important feature should be noted here. In a waveguide cavity used in the aforementioned experiments, the frequency range of the entire MDM spectrum lies above the cut-off-frequency of a dominant mode and below cut-off-frequencies of the high-order modes of the cavity [48]. It means that the *complete-set spectrum* of MDM oscillations in a ferrite disk is viewed as *virtual photons* in the cavity.

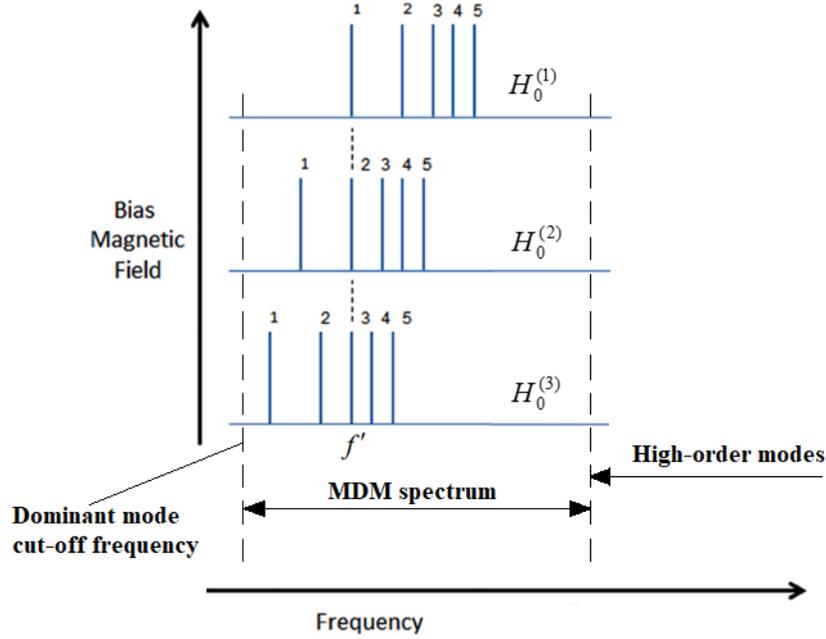

**Fig. 7.** Spectra of MDM resonances of a ferrite disk in a cavity. There is a clear correspondence between the frequency and the bias magnetic field. The entire MDM spectrum lies in the frequency region where the high-order cavity modes are virtual photons.

The microwave measurement reflects interaction between a microwave cavity and a MDM particle. If a MDM particle is under interaction with a "classical electrodynamics" object – the microwave cavity – the states of this classical object change. The character and value of these changes depend on the MDM quantized states and so can serve as its qualitative characteristics. It means that, in neglect of losses, there should exist a certain *uncertainty limit* stating that

$$\Delta f \, \Delta H_0 \geq \text{uncertainty limit} \tag{7}$$

This uncertainty limit is a constant which depends on the disk size parameters and the ferrite material property. Due to the uncertainty principle, virtual particles are considered as fluctuations in quantum fields that exist for a very narrow frequency deviation $\Delta f$ and very narrow region of a bias magnetic field $\Delta H_0$. Beyond the frames of the uncertainty limit, one has a continuum of energy.

Quantization of orbital angular momentums of dipolar magnons in a ferrite disk at a given signal frequency $\omega$ can be understood from the model represented in Fig. 8. For an observer located on a ferrite disk, the magnetization dynamics is seen as the Larmor-clock magnon dynamics. There are standing-wave oscillations of *G*-mode MS wave functions. The quantization of *G* modes is with numbers $q$ of radial variations and azimuth numbers $\nu$ [14, 19, 20, 28, 29]. Due to the cavity fields, there is an orbital rotation of *G*-mode magnons at frequency $\omega$. An observer located on a microwave cavity sees $2\omega$ rotations (*L*-mode magnon dynamics) of the wavefronts at the resonances. The *L*- mode magnetization dynamics of MS wave functions are accompanied by the electric polarization dynamics of ES wave functions.

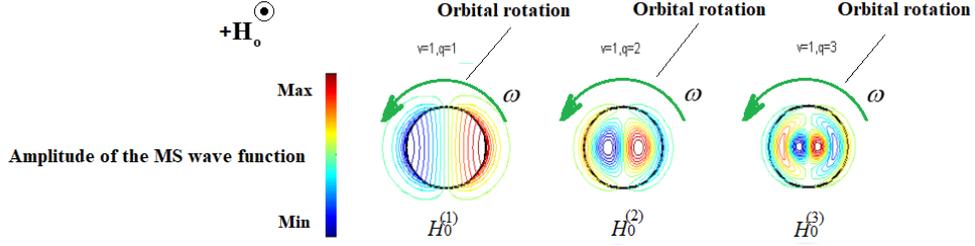

**Fig. 8.** Quantization of OAMs of dipolar magnons in a ferrite disk at a given signal frequency $\omega$. The scattered fields can absorb or emit the orbital-modulation quanta $\hbar\omega$. The observer on a ferrite disk can see that magnetization dynamics is Larmor-clock magnon dynamics. There are standing-wave oscillations of *G*-mode MS wave functions. The quantization of *G* modes is with numbers $q$ of radial variations and azimuth numbers $\nu$. Due to the cavity fields, there is an orbital rotation of *G*-mode magnons at frequency $\omega$. At MDM resonances, the observer on a microwave cavity sees $2\omega$ rotations (*L*-mode magnon dynamics) of the wavefronts.

The interaction of ferrite-disk MDM magnons with the cavity is via virtual photons. In this structure, the *PT* symmetry and angular momentum balance of the MDM resonances arise due to topological electric charges and currents induced on the metal walls of the cavity. The balance of angular momenta of a ferrite disk and cavity walls is related by biorthogonal relations. A vacuum zone at the vicinity of a ferrite disk is a region determined by a complete-set spectrum of quasistatic oscillations. Because of the superradiance effect [14], the energy outside this zone will be greater than inside, resulting in a torque that rotates the cavity fields. Due to the uncertainty principle (7), virtual particles are considered as fluctuations in quantum fields that exist for a very narrow frequency deviation and very narrow region of a bias magnetic field. These virtual particles are created in pairs of "particles" and "anti-particles", which are the right-handed and left-handed helical modes [14, 38].

The variation of a bias magnetic field at a constant frequency of the microwave signal leads to the appearance of quantized edge magnetic and electric currents on the lateral surface of the ferrite disk and quantized electric and magnetic charges on the ferrite-disk planes. The mesoscopic-scale quantization is considered as a confinement effect of the MS and ES oscillations [15, 19, 29, 43]. This occurs due to the ME effect. This model, explaining experiments in Refs. [39 – 42], is illustrated in Fig. 9. From an analysis of the reflection and transmission characteristics, shown in Figs. 9 (*a*), (*b*), one can see that the microwave energy accumulated in cavity sharply decreases at the MDM resonances. At the MDM resonance, the cavity extracts energy from an internal energy of a ferrite disk. It means that at this resonance, the cavity acquires *negative energy* relative to the microwave source feeding the cavity.

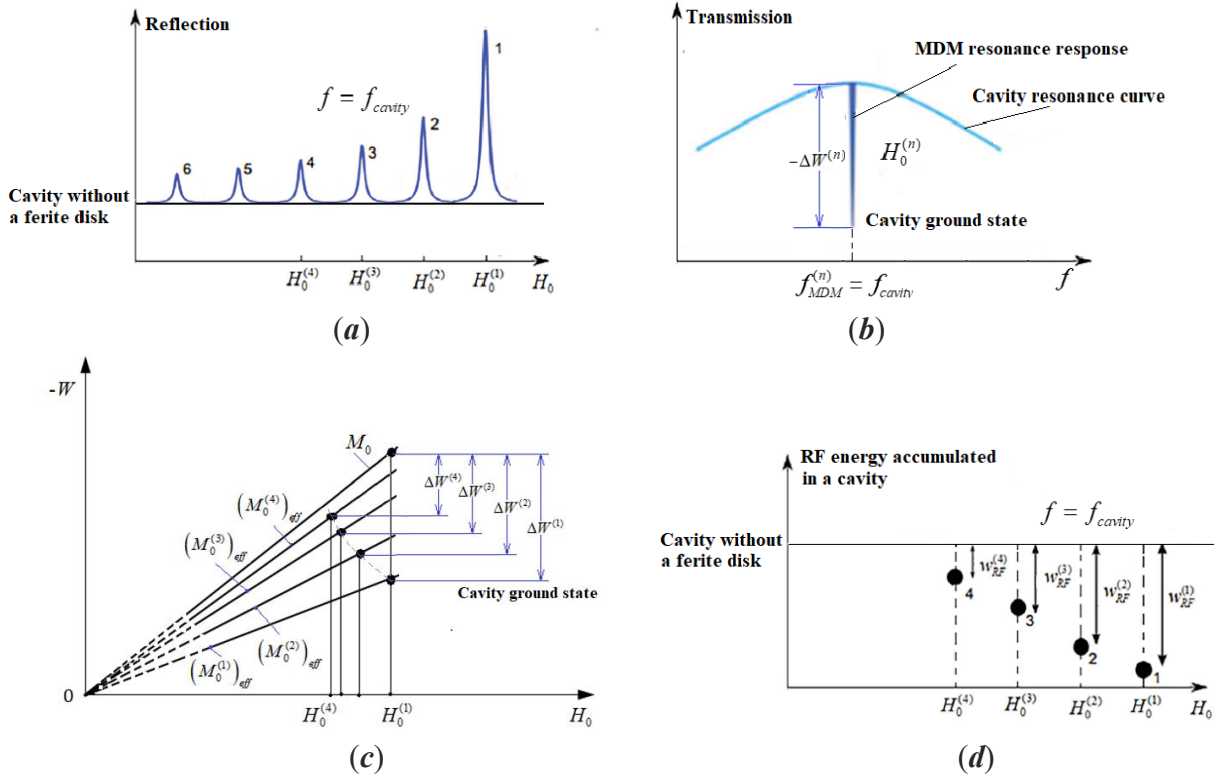

**Fig. 9.** At the MDM resonance, the cavity extracts energy from an internal energy of a ferrite disk. (*a*), (*b*) the reflection and transmission characteristics; (*c*) a model of discretization of magnetic energy in a ferrite disk at $\omega$ = const; (*d*) at the MDM resonances the cavity has negative-energy eigenstates.

Magnetic energy of a sample in an external (bias) magnetic field is determined by the demagnetization field. The demagnetization field is the magnetic field generated by the magnetization in a magnet and the demagnetization factor determines how a magnetic sample responds to the bias magnetic field. It is evident that for the MDM spectra, obtained at variation of a bias magnetic field $H_0$ and at a constant signal frequency $\omega$, a discrete reduction of magnetic energy of a ferrite disk at the MDM resonance should occur because of quantization of the demagnetization field. A model of discretization of magnetic energy in a ferrite disk at $\omega$ = const is shown in Fig. 9 (*c*). The slope of the straight lines is determined by DC magnetization. $M_0$ is saturation magnetization of a homogeneous ferrite material. At the MDM resonances, the slope of the straight lines is determined by $\left(M_0^{(n)}\right)_{eff}$.

$\Delta W^{(n)}$ shows the microwave energy extracted from a ferrite disk at the $n^{\text{th}}$ MDM resonance. Discretization of magnetic energy is shown for the first four MDMs. In Fig. 9 (*d*), we can see that at the MDM resonances the cavity has negative-energy eigenstates. It means that at these resonances the cavity returns discrete portions of the energy to the microwave source.

The thermodynamic state of a MDM ferrite disk can be changed only at the expense of effects on a microwave cavity. When a MDM resonator does *positive work* (leading to an increase in entropy), the field in the cavity does *negative work* (which leads to a decrease in entropy). Quasiparticles resulting from strong coupling of microwave photons with MDM resonators are called

ME polaritons. In Fig. 10, we show two counter propagating waves in a microwave waveguide at a certain time phase and a given direction of a bias magnetic field. The picture is obtained based on numerical results of Ref. [49]. The superposition of these waves gives a specific field structure in the cavity.

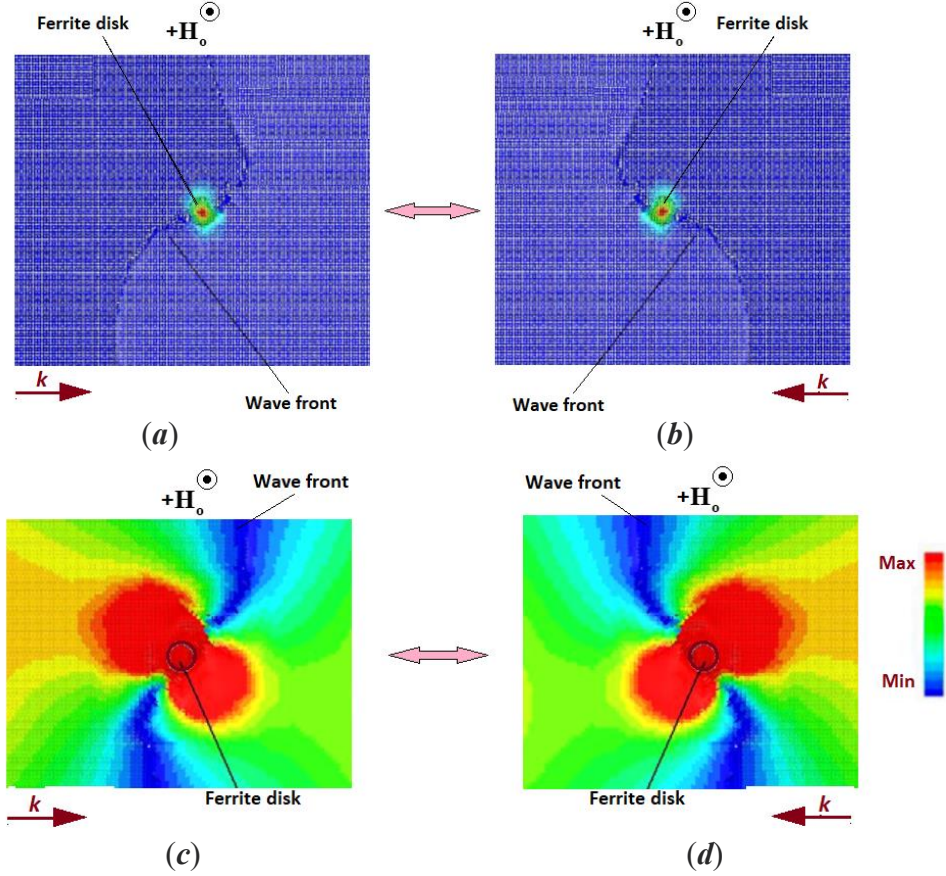

**Fig. 10.** ME polaritons in a cavity. Due to curved fronts, ME polaritons in a cavity have reduced energy.

When an incident EM wave scatters from an obstacle, it is partially reflected and partially transmitted. If the obstacle is rotating, waves can be amplified in the process, extracting energy from the scatterer. In our case, incident EM waves are amplified by extracting energy from the collective motion of orbitally rotating spinning magnetic dipoles. At the MDM resonances, transfer between angular momenta of the magnetic insulator and the microwave structure, demonstrates generation of vortex flows with fixed handedness. In our analysis, we emphasize that the quantized orbital angular momentums of dipolar magnons can be physically observed only if the Larmor-clock synchronization takes place. Are the Larmor-clock time and the EM-wave phase time the well-established times? With the probabilistic requirement of the uncertainty limit (7), we should have the average duration of the process as a real quantity.

In the cavity ME polaritons, the Larmor-clock synchronization can be clearly observed by spatial orientations of the electric field vectors. This is shown in Fig. 11. The effect of synchronization of the Larmor-clock arrows (originated from the magnetization dynamics in the disk) makes the *curved wavefront* to be thermodynamically favored. For a given signal frequency $\omega$, we have $2\omega$ OAM rotation in a ferrite disk. This means that the curved topological structure of the EM wavefront appears at the same form after $t = T/2$, where $T$ is a period of the RF radiation. In Fig. 12, we illustrate how the Larmor-clock arrows are synchronized with the radiative arrows on the EM-wave wave front. Fig. 12 (*a*) shows the structure of the wavefront in a certain region of space at time moment $t_1$. The picture is accompanied by the Larmor-clock arrows, located in a specific position. For the same wavefront structure at time moment $t_2 = t_1 + T/2$, shown in Fig. 12 (*b*), the direction Larmor-clock arrow rotates at $180°$.

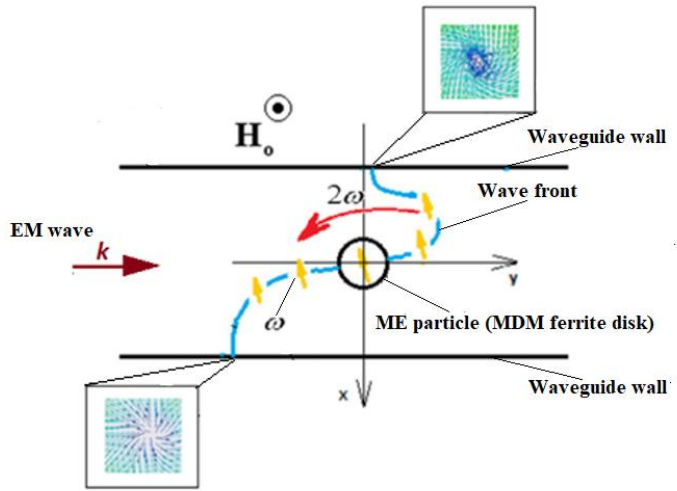

**Fig. 11.** Schematic representation of wavefronts and topological-phase electric fields at a given time phase. Yellow arrows show the spatial orientations of the electric field vectors. The directions of these arrows coincide with the directions of the Larmor-clock arrows. There is synchronization between the arrows of Larmor clock of magnetization dynamics in the disk and arrows of the "clock" of surface electric currents on the metal walls. The inserts show surface electric currents for the 1st MDM on the top and bottom walls of the waveguide [49].

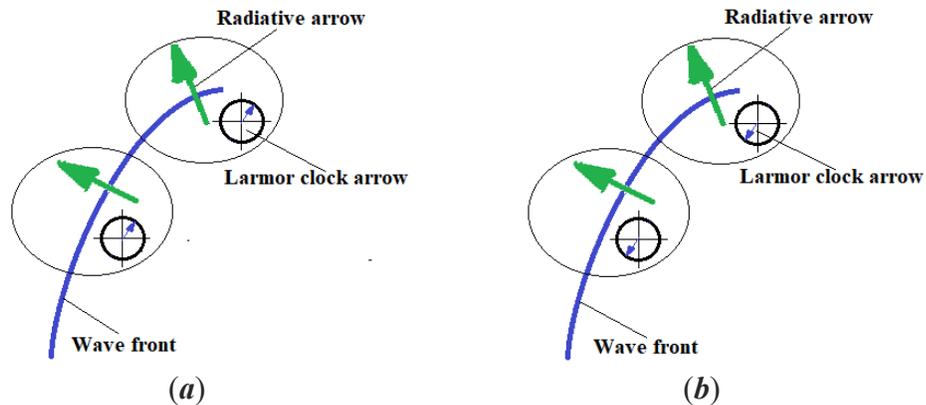

(*a*)          (*b*)

Fig. 12. The effect of synchronization between the Larmor-clock arrows (originated from the magnetization dynamics in the disk) and the radiative arrows of the EM-wave curved wavefront. (*a*) The structure in a certain region in a space at time moment $t_1$. (*b*) The structure at time moment $t_2 = t_1 + T/2$, where $T$ is a period of the RF radiation.

In the shown ME-polariton field structure, unilateral circulations of power flows occur in the cavity due to curved wave fronts. There are helical modes in vacuum. For each of these modes, originated from MDM resonances, we have specific topology. The directions of wave fronts of helical waves in a cavity are at certain correlation with the directions of the Larmor-clock arrows. Inside a ferrite disk, one observes helical-mode resonances of quasistatic oscillations (Fig. 13). The quasistatic wave functions are a four-component wave functions, or bispinors [14, 38].

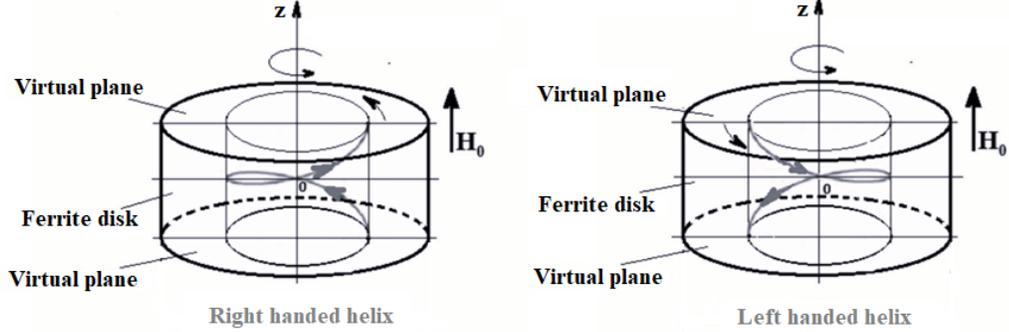

Fig. 13. The double-helix resonance inside a ferrite disk. The helices are displayed in such a way that the height of one complete helix turn is between two virtual planes. In the real structure of an open ferrite disk, the disk thickness is much less than the pitch.

## IV. CONCLUSION

We consider an intrinsic magnon spin-orbit coupling in a 3D-confined magnetic insulators due to the dipolar interaction. In the case of the MDM resonator, we have a chosen direction of time determined by the direction of a biased magnetic field. Transitions in time for magnetization dynamics are counted by a Larmor clock. For the $\theta = 2\pi$ orbital rotation, we cannot generate a *back-action*, necessary for any resonance (*normalized*) state. The back-action can be generated if we *additionally* use another disk which has the $\theta = 2\pi$ orbital rotation at an oppositely directed bias magnetic field. In such a *PT* symmetric system, MDM resonators, connected through long-range dipolar interaction, are in a *biorthogonality* relationship. The OAM balance occurs when both types of magnetic bias fields are used. In any separate disk, the resonant state occurs at the $4\pi$ orbital rotation. It means that the OAM of magnon the MDM resonator is $\mathcal{L} = \pm \frac{1}{2}\hbar$. Power-flow vortices of two ferrite-disk resonators are coupled due to long-range dipolar interaction. The thermodynamic state of one resonator can be changed only at the expense of effects on another resonator. To achieve thermodynamic equilibrium, the closed-loop unilateral wave propagations in the resonators must be mutually synchronized. This synchronization occurs due to Larmor clocks, operating in both resonators.

In microwave experiments with MDM oscillations, a *single* ferrite disk is used in a microwave structure. To understand the underlying mechanism of angular momentum balance in such a single-disk system, it is necessary to analyze how a MDM ferrite disk interacts with metal walls of the microwave waveguide. In the dynamic processes in these magnetic insulators, it is necessary to take into account not only MS interactions, but also ES interactions. The mesoscopic-scale quantization is described based on the MS and ES wave functions. Such quantized states are well observed in the cavity ME effect. Due to the topological action of the azimuthally unidirectional transport of energy in a MDM-resonance ferrite sample there exists the opposite topological reaction on a metal screen placed near this sample. This is a topological Lenz's effect. Rotation of the fields in a ferrite disk is in the angular momentum balance with the rotation of electric currents and topological charges induced on the metal walls.

The interaction of cavity photons with the MDM ferrite disk is designated as ME cavity polaritons. The entire spectrum of MDM resonances lies in the frequency region where the high-order cavity modes are virtual photons. Due to the uncertainty principle, virtual particles are considered as fluctuations in quantum fields that exist for a very narrow frequency deviation and very narrow region of a bias magnetic field. Beyond the frames of the uncertainty limit, one has a continuum of energy. While keeping the solid body motionless, the rotating fields of MDM resonators modify the momentum of quantum vacuum. The mesoscopic-scale quantization is considered as a confinement effect of the MS and ES oscillations. This occurs due to the ME effect. At the MDM resonance, the cavity extracts energy from an internal energy of a ferrite disk. It means that at this resonance, the cavity acquires negative energy relative to the microwave source feeding the cavity. Incident EM waves are amplified by extracting energy from the collective motion of orbitally rotating spinning magnetic dipoles. In our analysis, we emphasize that the quantized orbital angular momentums of dipolar magnons can be physically observed only if the Larmor-clock synchronization takes place. The effect of synchronization of the Larmor-clock arrows (originated from the magnetization dynamics in the disk) makes the curved wavefront to be thermodynamically favored.

We consider an arrow of time in relation to the Larmor clock. The Larmor-clock mechanism is pure quasistatic. It means that it extends to infinity. Due to the Larmor-clock mechanism, we can observe the one-way closed-loop circulation of the arrow of time in a microwave cavity. This circulation of radiative arrows involves the power-flow propagation in a curved space-time. While for an incident EM wave there is no difference between left and right, in the fields scattered by a ME particle one should distinguish left from right. There are helical modes in vacuum. Inside a ferrite disk, one observes helical-mode resonances of quasistatic oscillations.